\UseRawInputEncoding
\documentclass{cimento}

\usepackage{bm}

\usepackage{graphicx}  

\def\ltap{\raisebox{-.6ex}{\rlap{$\,\sim\,$}} \raisebox{.4ex}{$\,<\,$}} 
\def\gtap{\raisebox{-.6ex}{\rlap{$\,\sim\,$}} \raisebox{.4ex}{$\,>\,$}} 

\title{Multi-jet physics  at high-energy colliders and \ETC \ TMD parton evolution 
}
\author{A.~Bermudez Martinez\from{ins:x}, 
F.~Hautmann\from{ins:y}
        \atque
M.L.~Mangano\from{ins:z}
}
\instlist{\inst{ins:x} DESY, Hamburg
  \inst{ins:y} University of Oxford / University of Antwerp  
   \inst{ins:z} CERN, Geneva}

\begin{document}

\maketitle

\begin{abstract}
We discuss  implications of  the evolution of   
transverse momentum dependent (TMD) parton distributions on the structure of multi-jet states at high energies.~In particular 
we analyze  the theoretical systematics associated with multi-jet merging.  We   
 introduce a new merging methodology incorporating TMDs,   illustrate its main features and present a  comparison of   
our theoretical results with 
experimental measurements for $Z$-boson + jets production at the Large Hadron Collider (LHC).\\  
\vskip 0.1 cm 
\hskip 0.1 cm \em{Talk given by F.~Hautmann at the Rencontres de 
Physique de la Vall{\' e}e d'Aoste\\ 
\hspace*{4.2cm} La Thuile, March 2021}
\end{abstract}

\section{Introduction}

In the last few years, experimental studies of  Drell-Yan (DY)  lepton pair 
production~\cite{Aad:2015auj,CMS:2016mwa,ATLAS:2019zci,CMS:2019raw} 
and deep inelastic scattering~\cite{Agostini:2020fmq,Aidala:2020eah}  
have underlined the role of TMD parton evolution~\cite{Angeles-Martinez:2015sea}.   See 
e.g.~\cite{Hautmann:2020cyp} for a recent study   of the interplay 
between perturbative and non-perturbative effects induced  by TMD evolution in the 
DY 
spectrum at low transverse momenta.   

The impact of TMD distributions on the high transverse momentum region and on 
multi-jet production, on the other hand, is as yet unexplored, and constitutes 
the subject of the work presented in this article.    

Theoretical predictions for multi-jet observables have relied for the past twenty years on 
``merging''  
techniques~\cite{Catani:2001cc,Lonnblad:2001iq,Mangano:2002,Mrenna:2003if,Alwall:2007fs}  to 
combine matrix-element and parton-shower event generators. 
The former 
describe the underlying hard process with bare partons providing the primary sources for widely separated jets;  the latter 
describe the evolution of partons by radiative processes predominantly at small angles; and the two are 
sewn together,  so as to avoid either double counting or missing events, via a ``merging scheme''  and 
merging scale. 
The choice of the merging scheme and merging scale is one of the main theoretical systematics 
in studies of multi-jet final states at high-energy hadron colliders,  investigated at  
 leading order (LO)~\cite{Alwall:2007fs,Mangano:2006rw} 
and next-to-leading order (NLO)~\cite{Frederix:2012ps,Hoeche:2012yf,Lonnblad:2012ix,Bellm:2017ktr}.  

Transverse momentum recoils in the   shower  evolution may be taken into account through 
TMD parton distributions~\cite{Angeles-Martinez:2015sea} and 
can influence  the theoretical systematics associated with combining 
matrix-element and parton-shower contributions~\cite{Dooling:2012uw,Hautmann:2013fla},   
  thus affecting  the dependence of multi-jet cross sections on the merging scale. 
  Motivated by this observation, 
  in~\cite{Martinez:2021chk} we devise a systematic  procedure of multi-jet merging, dubbed 
  ``TMD merging",  which 
  extends  to the case of TMD parton evolution the familiar merging 
approach~\cite{Catani:2001cc,Lonnblad:2001iq,Mangano:2002,Mrenna:2003if,Alwall:2007fs}.  
We use TMD merging  to analyze theoretical systematic uncertainties in multi-jet observables  
and 
to 
 perform comparisons of theoretical predictions with  experimental measurements for 
 $Z$-boson + jets production at the LHC~\cite{Aaboud:2017hbk,Sirunyan:2018cpw}. 

The analysis~\cite{Martinez:2021chk} employs 
 the parton branching (PB) formulation of TMD evolution set out in~\cite{Hautmann:2017xtx}.  
It constructs a merging at LO level expanding on  
the MLM matching prescription~\cite{Mangano:2002,Mrenna:2003if,Alwall:2007fs,Mangano:2006rw}. 
A similar construction is possible starting from other approaches, such as CKKW-L~\cite{Catani:2001cc,Lonnblad:2001iq}. 

In what follows we begin by discussing parton  $k_T$ broadening effects due to TMD evolution,  and their  implications 
for multi-jet production (Sec.~2). Then we present the TMD merging method (Sec.~3),  
 illustrate a few  applications to  $Z$-boson + jets production (Sec.~4), and discuss the 
associated theoretical systematics (Sec.~5).   We finally give concluding remarks (Sec.~6).

\section{$k_T$ broadening from TMD evolution} 

We consider the broadening in the transverse momentum $k_T$ of the partonic initial state 
which results from  TMD evolution~\cite{Martinez:2021chk}. 
For a multi-jet final state characterized by the hard momentum-transfer scale $\mu$, we 
analyze the contribution to the production of an extra jet with transverse momentum 
$p_T <  \mu$ from  the high-$k_T$ tail of the initial state parton distribution, $k_T \gtap p_T$. 
To estimate this,  we introduce  integral TMD distributions $a_j$, obtained from  the TMD distributions 
${A}_j$  by $k_T$-integration as follows 
 \begin{equation} 
\label{eq2integraltmd}
a_j ( x, {\bm k}^2, \mu^2) = 
\int    { {d^2 {\bm k }^\prime} \over \pi} \  {\cal A}_j ( x , {\bm k }^{\prime 2} , \mu^2) \  \Theta ( {\bm k }^{\prime 2}-{\bm k }^{ 2} ) 
\; . 
\end{equation} 

The distribution $a_j$ evaluated at $k_T = 0$ gives the fully integrated initial-state distribution. 
We are interested in the fractional contribution to it from the tail arising above transverse momentum $k_T$, with $k_T$ of the order of 
the jet $p_T$. For any flavor $j$ we thus construct the ratio 
\begin{equation}
\label{normalized-itmd} 
R_j ( x, {\bm k}^2, \mu^2) = a_j ( x, {\bm k}^2, \mu^2) / a_j ( x, { 0}, \mu^2)  \; .  
\end{equation} 

\begin{figure}[hbtp]
  \begin{center}
	\includegraphics[width=.60\textwidth]{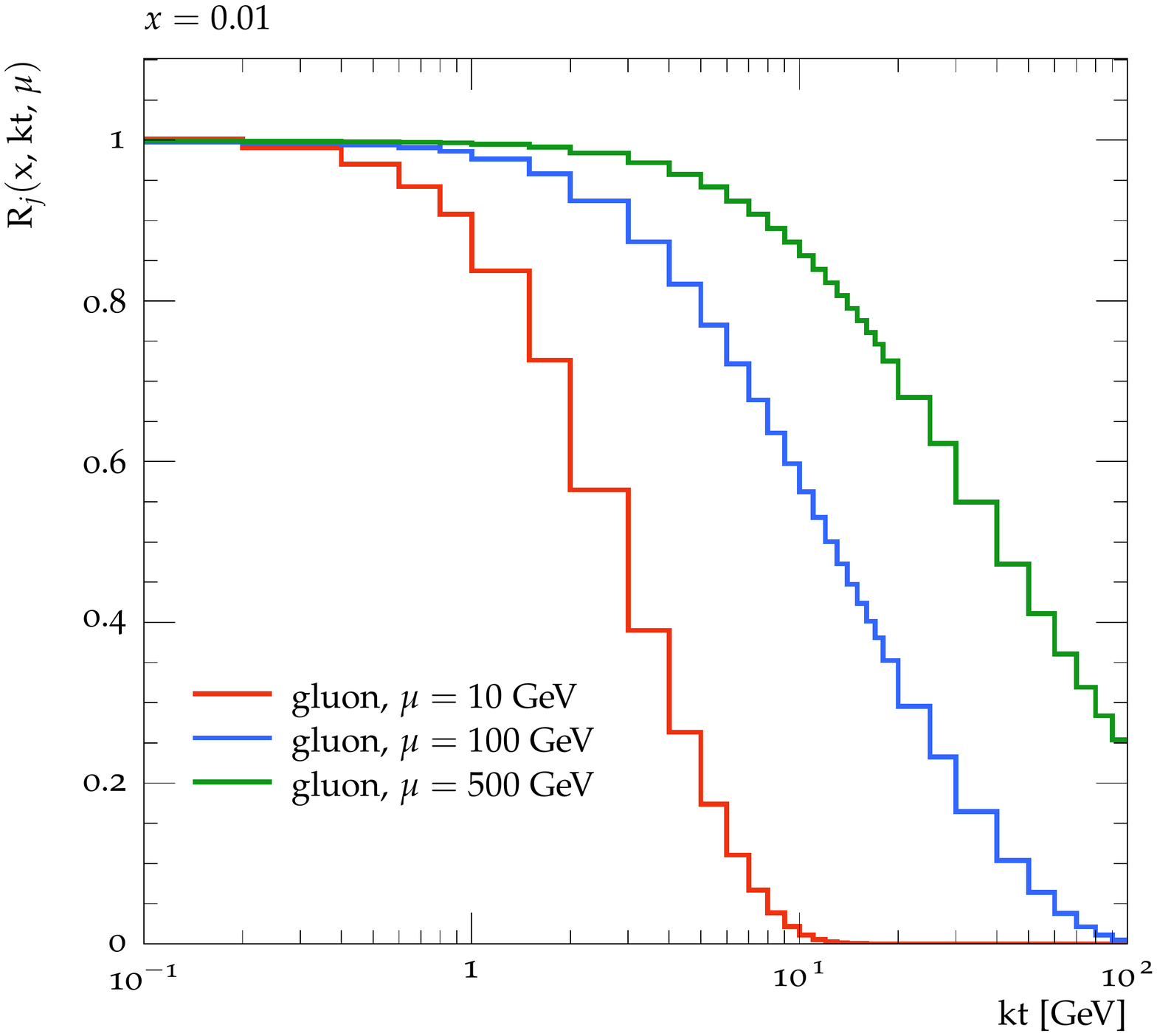}
  \caption{The $k_T$ spectrum of the integral  
  TMD  gluon distribution, normalized to $k_T=0$  as   in Eq.~(\ref{normalized-itmd}),  
  for longitudinal momentum fraction $x= 10^{-2}$ and different values of the evolution scale $\mu$. 
  The PB TMD Set~2~\cite{Martinez:2018jxt} is used.}
\label{fig0:iTMD-vs-kTmin}
\end{center}
\end{figure}

In Fig.~\ref{fig0:iTMD-vs-kTmin} we illustrate the $k_T$ dependence of Eq.~(\ref{normalized-itmd})   
by an example showing the  integral TMD gluon distribution $a_g ( x, {\bm k}^2, \mu^2) $ normalized to    $k_T=0$ for $x=10^{-2}$ and various values of $\mu$, 
 obtained from the TMD fitted in~\cite{Martinez:2018jxt}  to precision DIS data  using   \verb+xFitter+~\cite{Alekhin:2014irh} 
(for other available TMD fits, see the library~\cite{Abdulov:2021ivr}). We observe, for 
 instance, that for $\mu=100$ (500) GeV, there is a 30\% probability that the gluon has developed a transverse momentum larger than 20 (80) GeV.

While the distribution 
is falling off at large $k_T$, we find that for the jet transverse scales observed at the LHC  the contribution from the 
region $p_T \ltap k_T \ltap \mu$ is non-negligible when compared  
to that of an extra parton perturbatively emitted through hard-scattering matrix elements.
As a result, a merging methodology  is needed to avoid the double counting between 
the extra jet emission induced by the TMD initial-state evolution and that arising from the inclusion of a higher-order matrix element.
Such a methodology is  developed in~\cite{Martinez:2021chk,prepa21}, and is discussed in the next section.

\section{TMD multi-jet merging and differential jet rates}

Current multi-jet merging approaches provide techniques to combine  
samples of different parton multiplicity showered through emissions in the collinear 
approximation~\cite{Catani:2001cc,Lonnblad:2001iq,Mangano:2002,Mrenna:2003if,Alwall:2007fs}. 
The TMD merging approach~\cite{Martinez:2021chk} complements these approaches with the use 
of the TMD parton branching for the initial state evolution. 

The distinctive  features of TMD merging, compared to collinear merging, are embodied 
in three steps:  i) for any $n$-jet parton level event, initial-state transverse momenta $k_{T i}$ are generated 
according to the TMD distributions obtained as solutions of the PB evolution equations~\cite{Hautmann:2017xtx,Hautmann:2019biw}, 
but rejecting, owing to Sudakov suppression, $k_{T i} > \mu_{min}$, where   $\mu_{min}$ is the minimum energy 
scale in the $n$-jet hard sample; ii) initial state partons of the generated events are showered using the 
backward space-like shower evolution driven by the  PB  equations~\cite{Hautmann:2017xtx,Hautmann:2019biw}, while final state partons are 
showered using standard time-like showers; iii)   a merging prescription, such as 
MLM~\cite{Alwall:2007fs,Mangano:2006rw}, is applied between the  showered event and the event generated in i) 
including the $k_{T }$ boost. 
As noted earlier, one may construct a similar procedure by using 
prescriptions other than MLM, for instance CKKW-L~\cite{Catani:2001cc,Lonnblad:2001iq}. 

\begin{figure}[hbtp]
  \begin{center}
	\includegraphics[width=.49\textwidth]{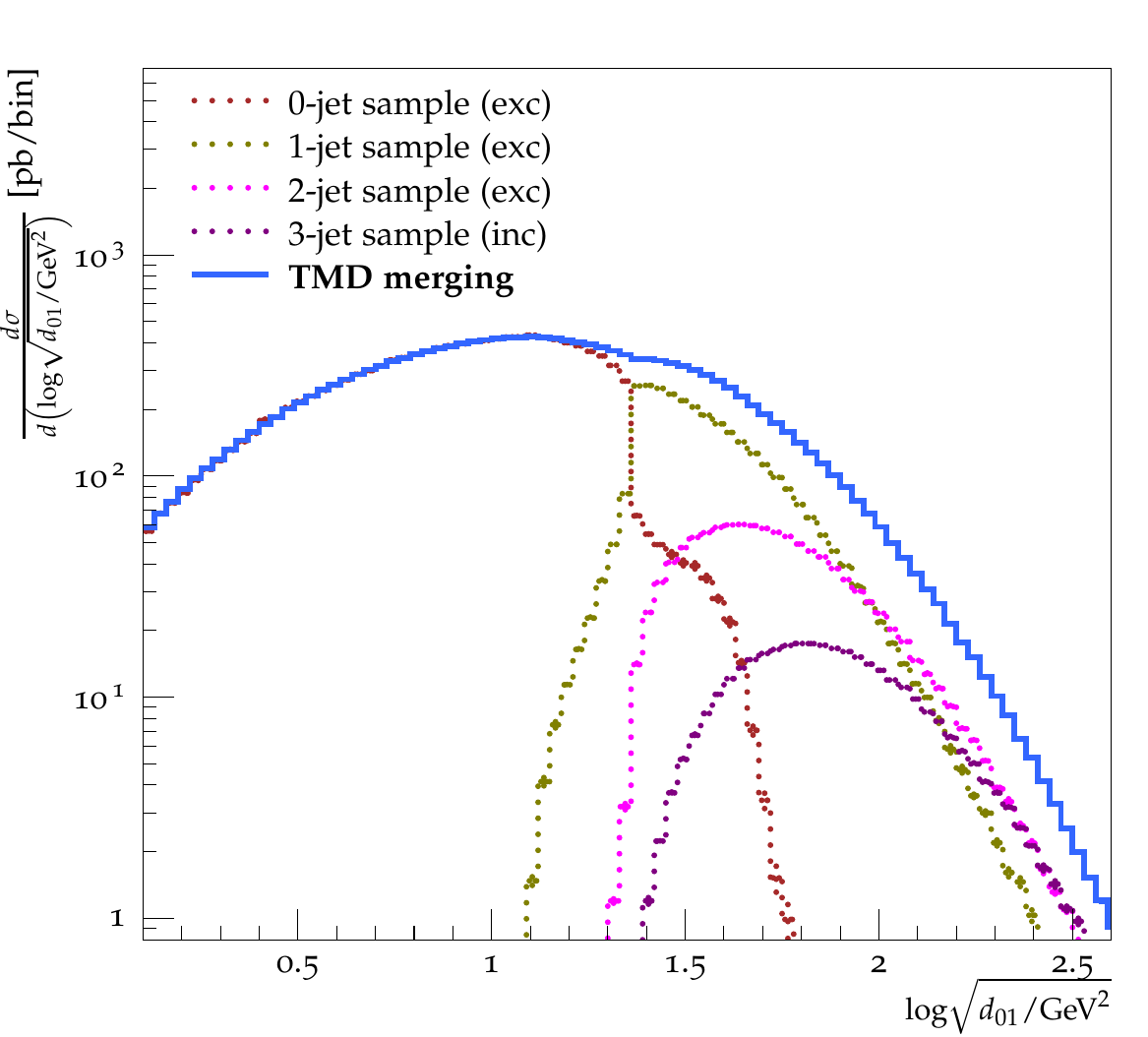}
	\includegraphics[width=.49\textwidth]{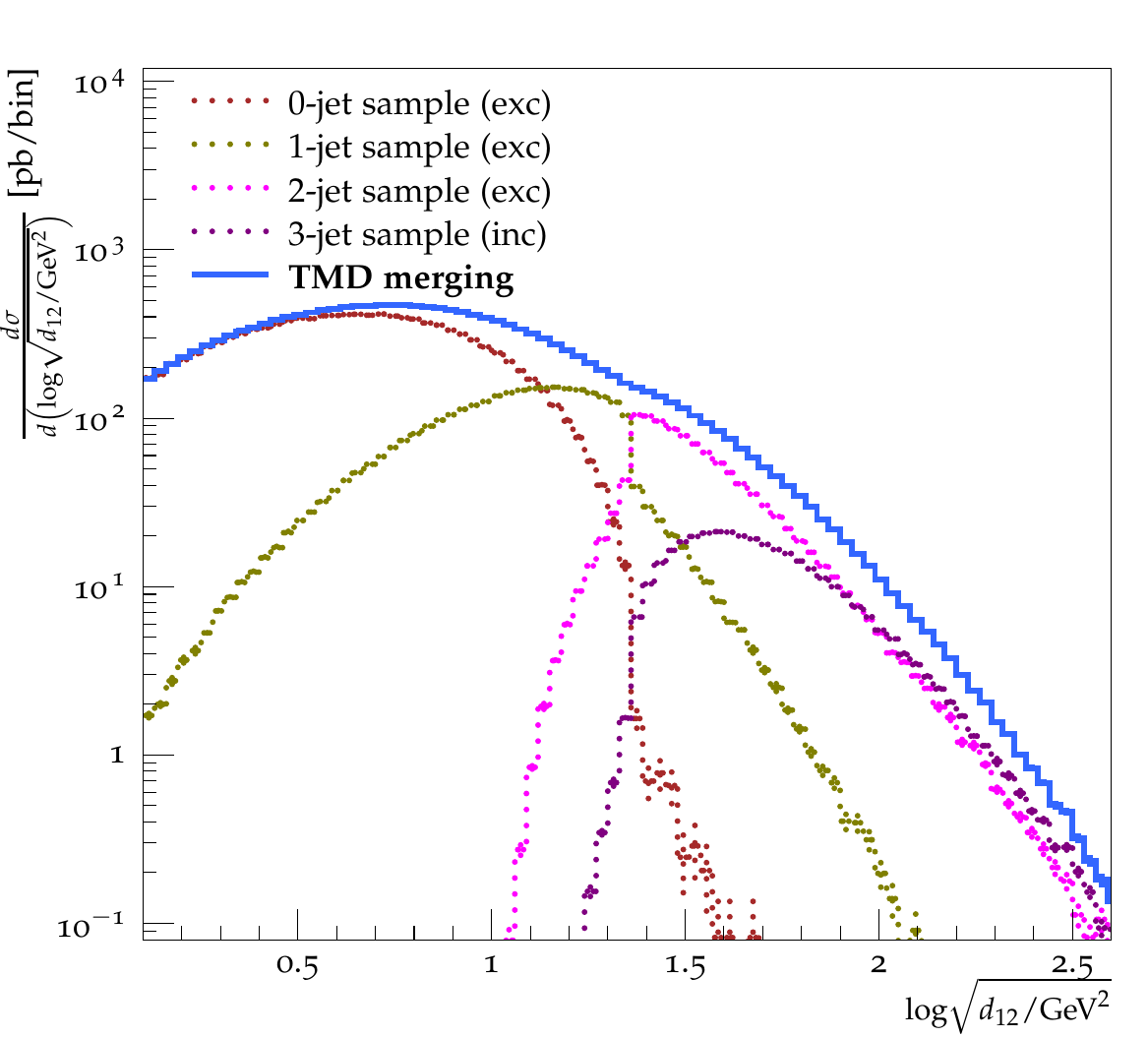}
    \includegraphics[width=.49\textwidth]{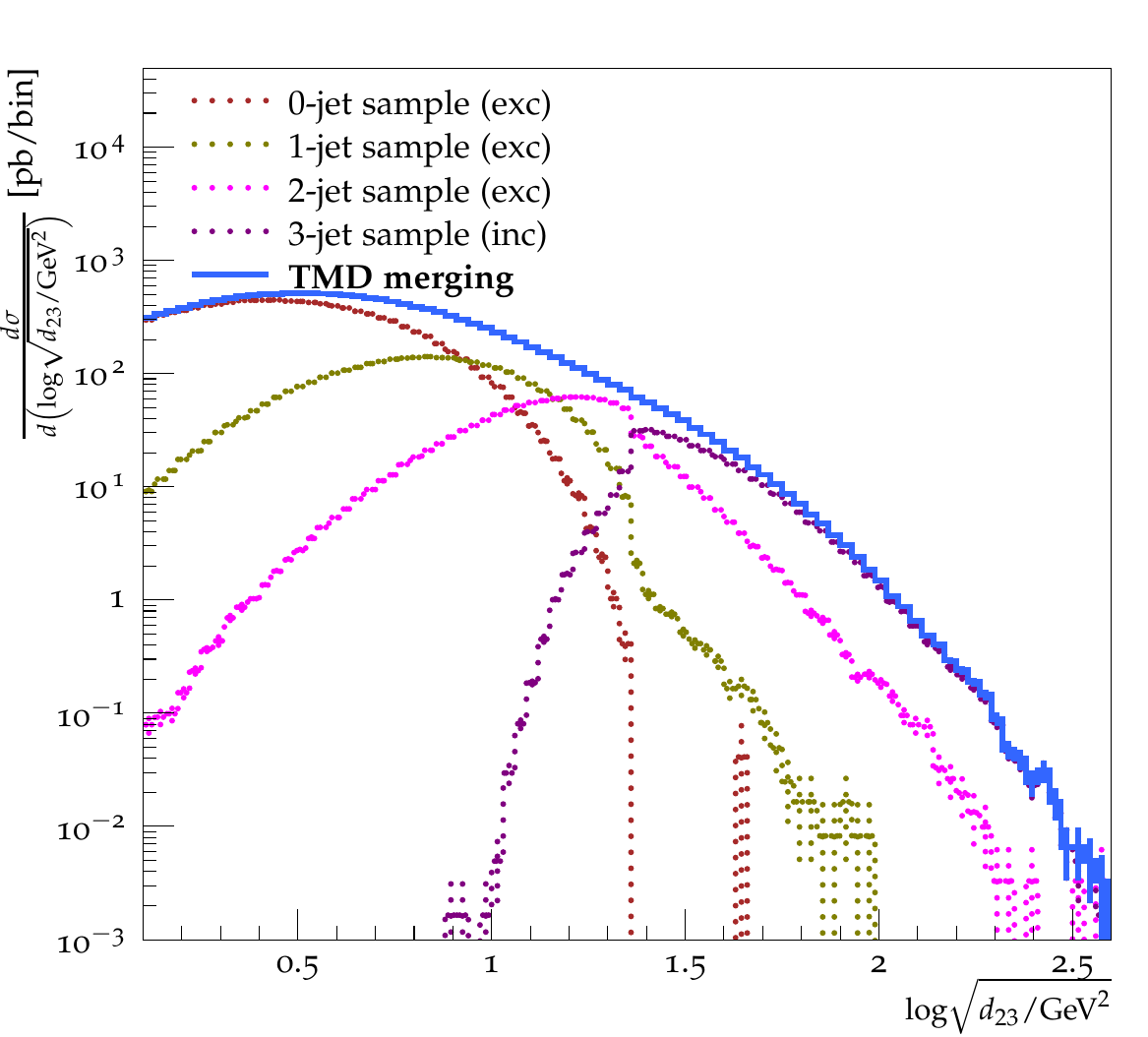} 
  \caption{The $d_{n,n+1}$  spectra for $n=0,1,2$  at parton level, where $d_{n,n+1}$ represents the energy-square 
  scale at which an $(n+1)$-jet event is resolved as an $n$-jet event in the 
  k$_\perp$ jet-clustering algorithm. The dotted curves represent the  contributions of the single-multiplicity samples while the solid curve corresponds to  their sum. 
  For each panel all jet multiplicities are obtained in exclusive (exc) mode except for the highest multiplicity which is calculated in inclusive (inc) mode.}
  \label{fig1-djr}
  \end{center} 
\end{figure}

We next illustrate the TMD merging methodology by computing the 
 differential jet rates (DJR) $d_{n,n+1}$ at parton level which result from the  k$_t$  jet clustering~\cite{Catani:1993hr} 
applied to final states containing a 
$Z$-boson. The $d_{n,n+1}$  represents the square of the energy scale at which an $n$-jet event is resolved as an $(n+1)$-jet event. 
Since the DJRs provide the splitting scales in the 
jet  clustering algorithm, they follow closely the measure used in the definition of the merging scale. Therefore, they are a 
powerful means to test the merging algorithm defined above. 

To do this calculation, we use 
 {\sc MadGraph5\_aMC@NLO}~\cite{Alwall:2014hca} to generate $Z$ $+0,1,2,3$ jet samples at LO with a 
 generation cut $q_{cut}=15$ GeV in $pp$ collisions at a center-of-mass energy $\sqrt{s} =$  8 TeV. 
 We use the event generator  {\sc Cascade}~\cite{Baranov:2021uol} to generate the TMD backward shower, and 
 {\sc Pythia}6.4~\cite{Sjostrand:2006za} for the final-state shower. 
We  apply the  parton distributions obtained from DIS fits in~\cite{Martinez:2018jxt} with $\alpha_s(M_Z) = 0.118$. 
The nominal value for the merging scale is chosen to be $\mu_m = 23$ GeV.

The results for the DJRs 
are shown in Fig.~\ref{fig1-djr}.  The dotted curves represent the $n$-jet sample contributions while the solid curve corresponds to their sum.   
All the multiplicities are calculated in exclusive mode except for the highest multiplicity which is calculated in inclusive mode.
 A clear separation between the different jet samples is seen at the merging scale value while the resulting overall prediction remains smooth.

\begin{figure}[hbtp]
  \begin{center}
	\includegraphics[width=.49\textwidth]{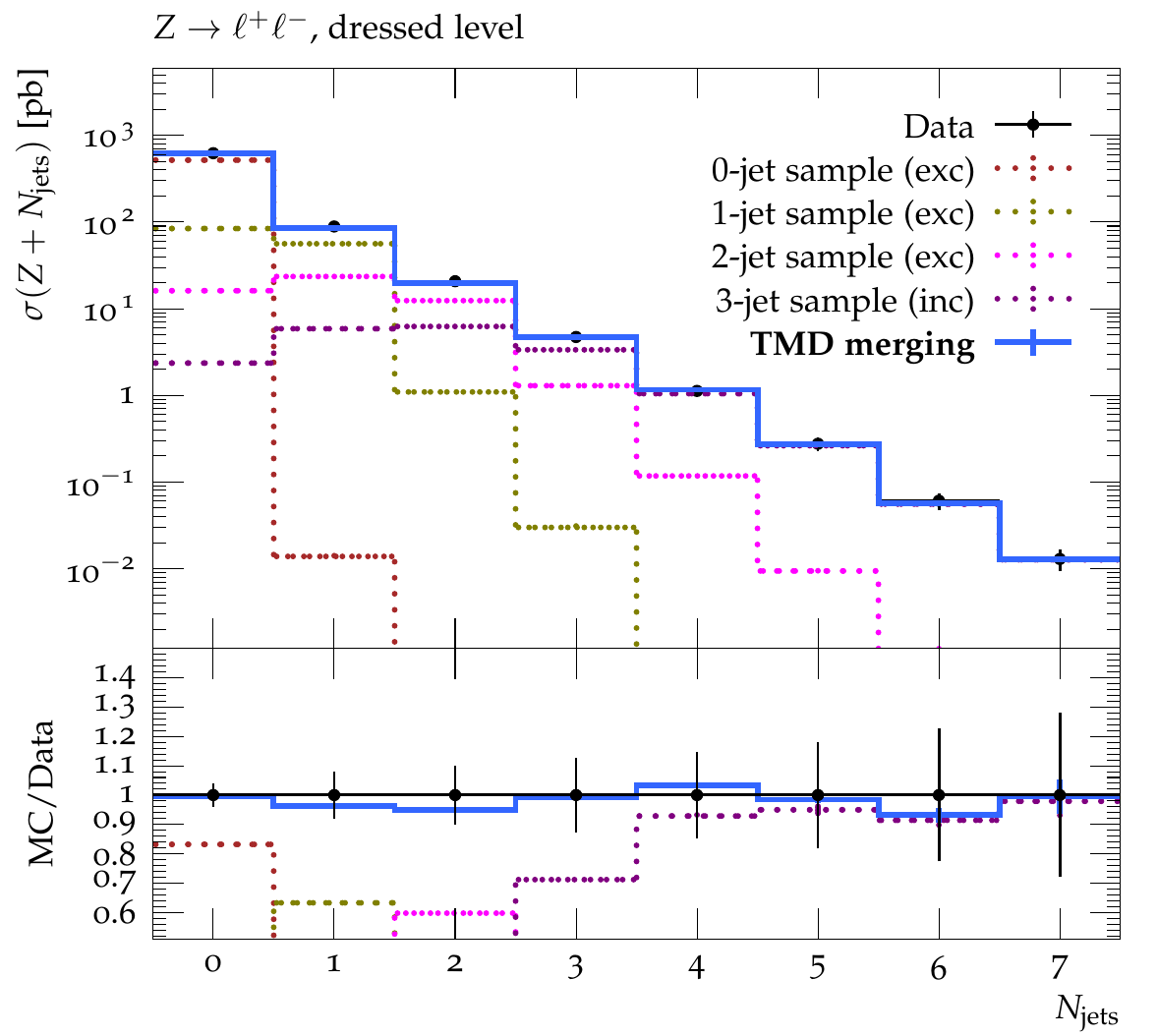}
	\includegraphics[width=.49\textwidth]{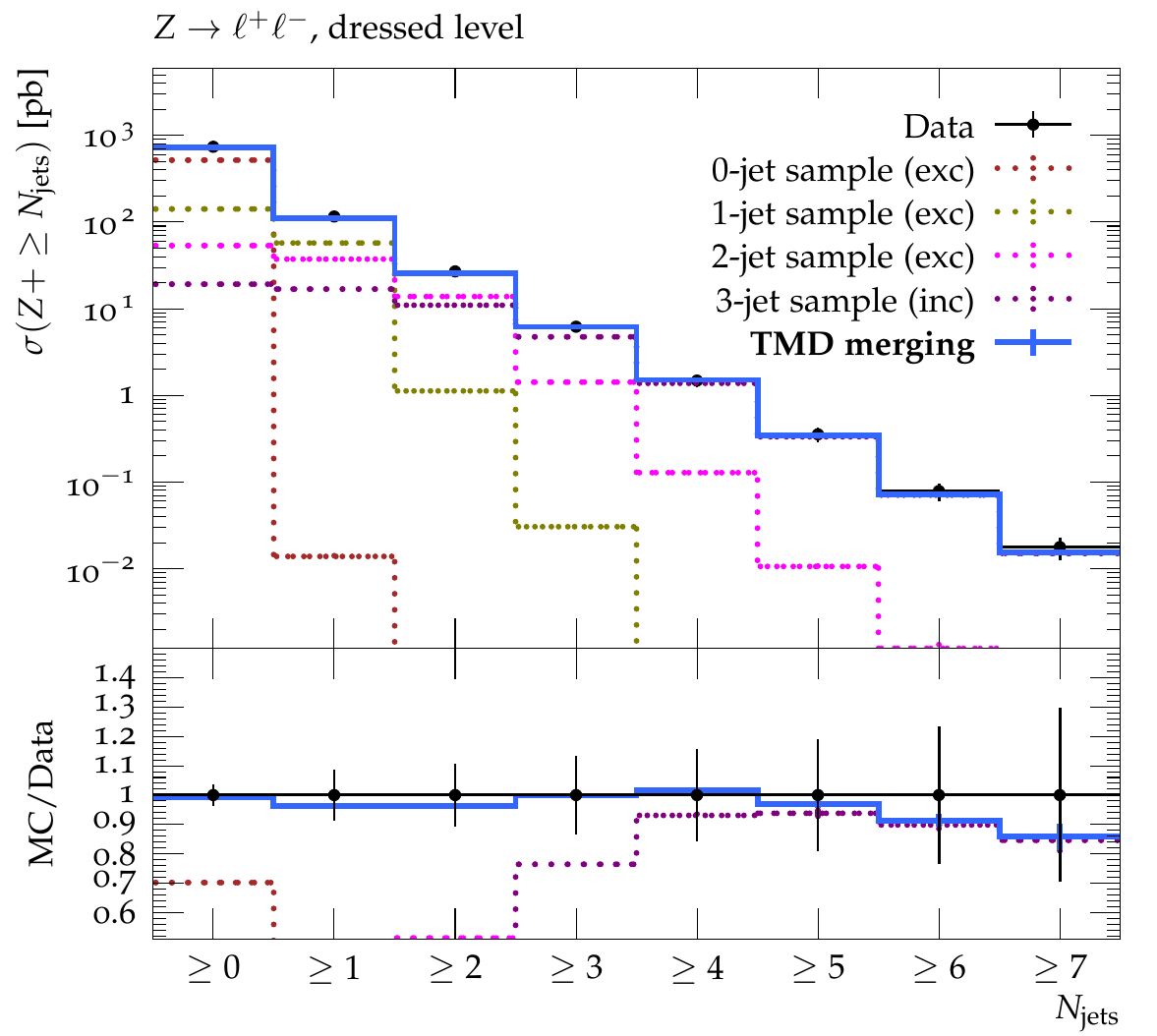}
  \caption{Exclusive (left) and inclusive (right) jet multiplicity distributions in the production of a $Z$-boson in association with jets. Experimental measurements by 
  ATLAS~\protect\cite{Aaboud:2017hbk} at $\sqrt{s} = 13$ TeV are compared to predictions using the TMD merging calculation. 
  Separate contributions from the different jet samples are shown. All the jet multiplicities are obtained in exclusive (exc) mode except for the highest multiplicity which is calculated in inclusive (inc) mode.}
  \label{fig2-multiplicities}
  \end{center}
\end{figure}

\section{$Z$-boson + jets production at the LHC: a case study}

In this section we present a few first applications of TMD merging to final states in DY production at the LHC.

We have first tested the method by evaluating the $Z$-boson transverse momentum spectrum. The 
result has been presented in~\cite{Martinez:2021chk}, and  compared with the measurements~\cite{Aad:2015auj}.  
The merged prediction is found to provide a good description of the data 
throughout the whole $Z$-boson $p_T$ spectrum, with the $Z+0$ jet sample constituting the
main contribution at low transverse momentum $p_T$ and the impact of larger jet multiplicities gradually increasing 
with increasing $p_T$. Thus the merged prediction~\cite{Martinez:2021chk}  retains the good description of the low-$p_T$ region 
already obtained in~\cite{Martinez:2019mwt} by applying TMD evolution, and improves the behavior in the 
high-$p_T$ region by merging TMD showers  with  higher  multiplicities.

Next, we consider jet observables measured in association with $Z$-boson production, and  compare the predictions 
with the ATLAS measurements~\cite{Aaboud:2017hbk}.  
In Fig.~\ref{fig2-multiplicities} we show the results for the exclusive (left) and inclusive (right) jet multiplicities 
in $Z$+jets events in $pp$ collisions at $\sqrt{s} = 13$ TeV. 
Jets are defined by the anti-$k_t$ algorithm~\cite{Cacciari:2008gp} with radius 
$R=0.4$, and are required to have $p_T>30$~GeV and $\vert \eta\vert<2.5$. 
The analysis is performed using {\sc Rivet}~\cite{Buckley:2010ar}.

\begin{figure}[hbtp]
  \begin{center}
	\includegraphics[width=.49\textwidth]{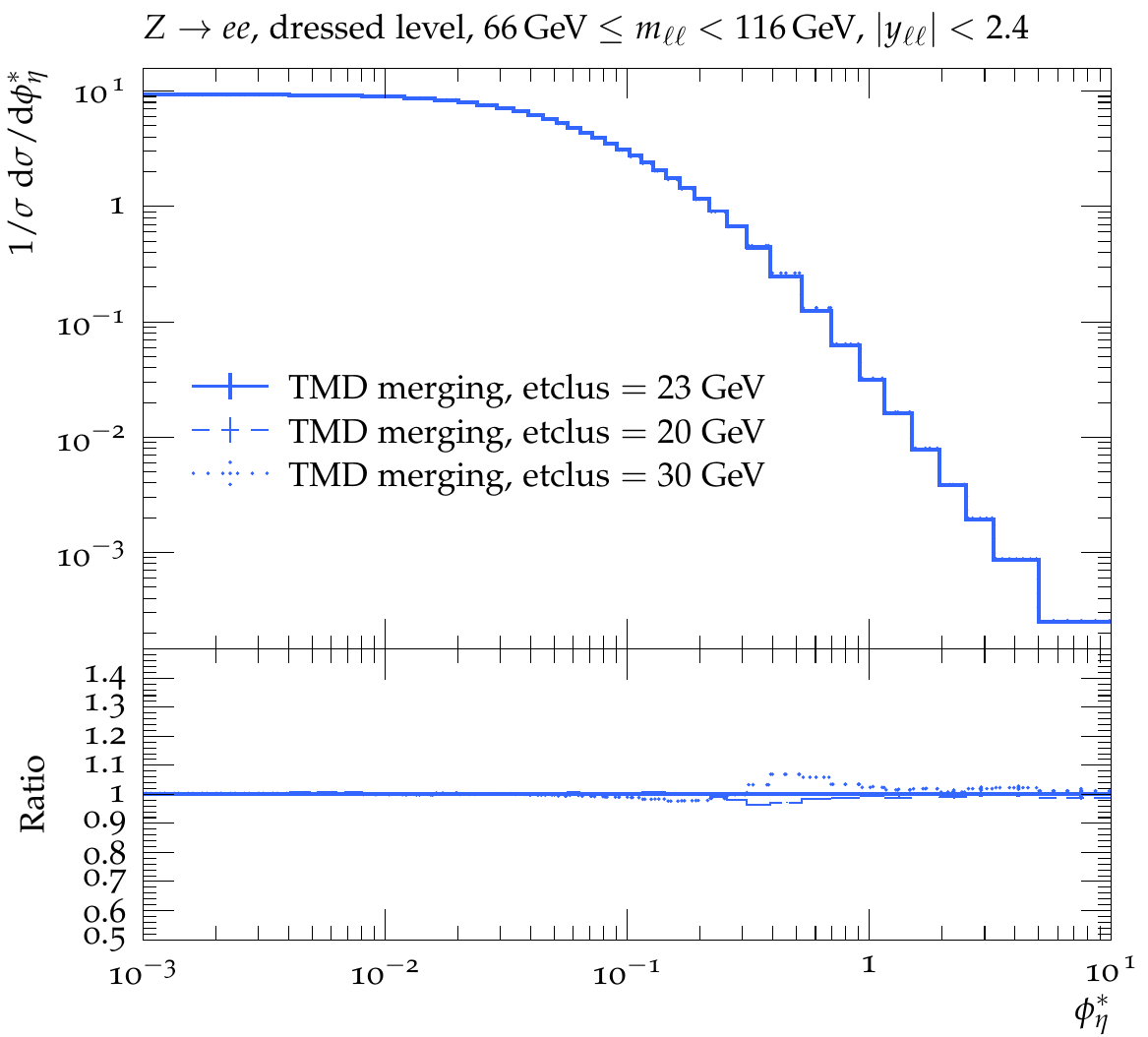}
	\includegraphics[width=.49\textwidth]{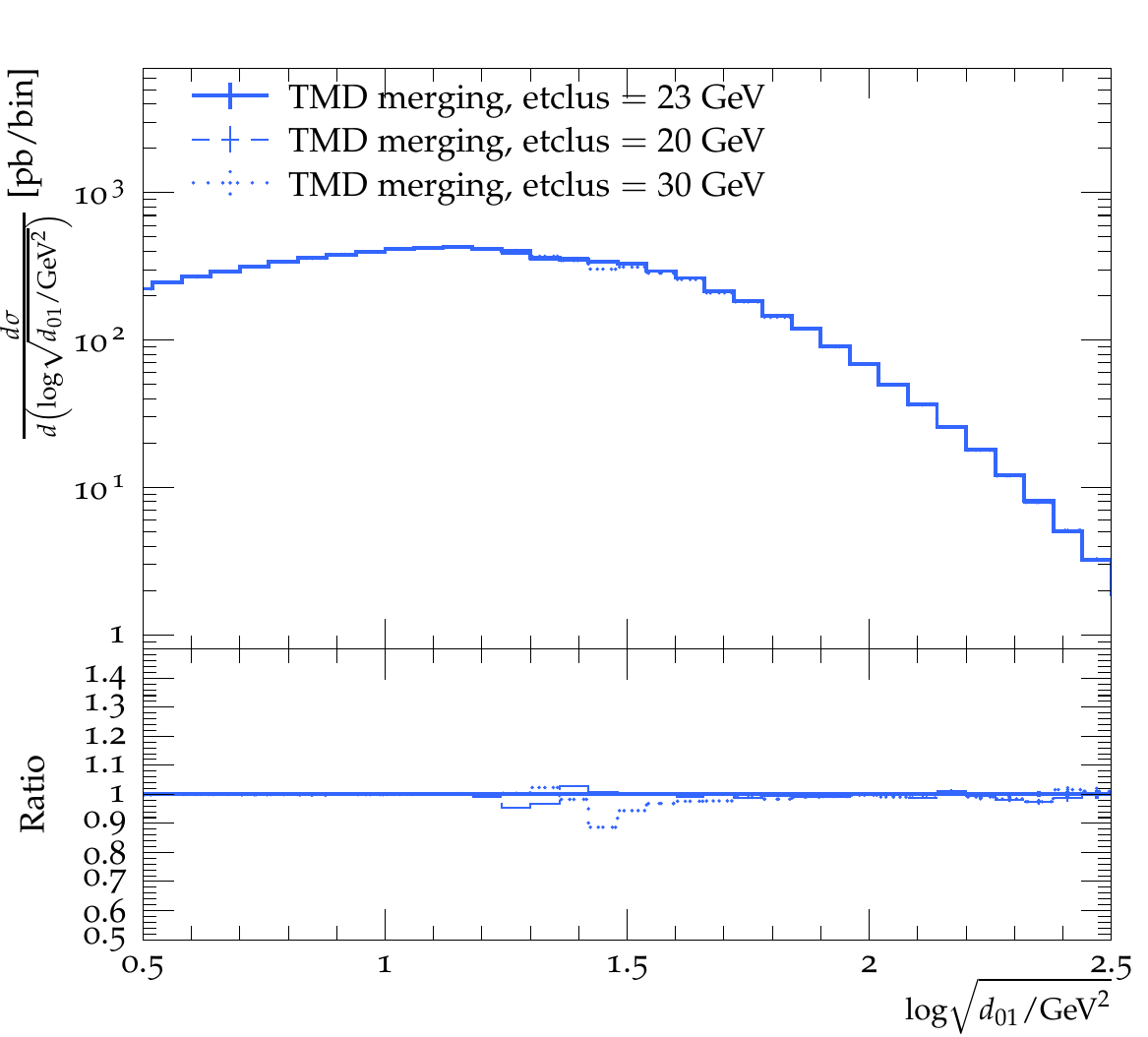}
    \includegraphics[width=.49\textwidth]{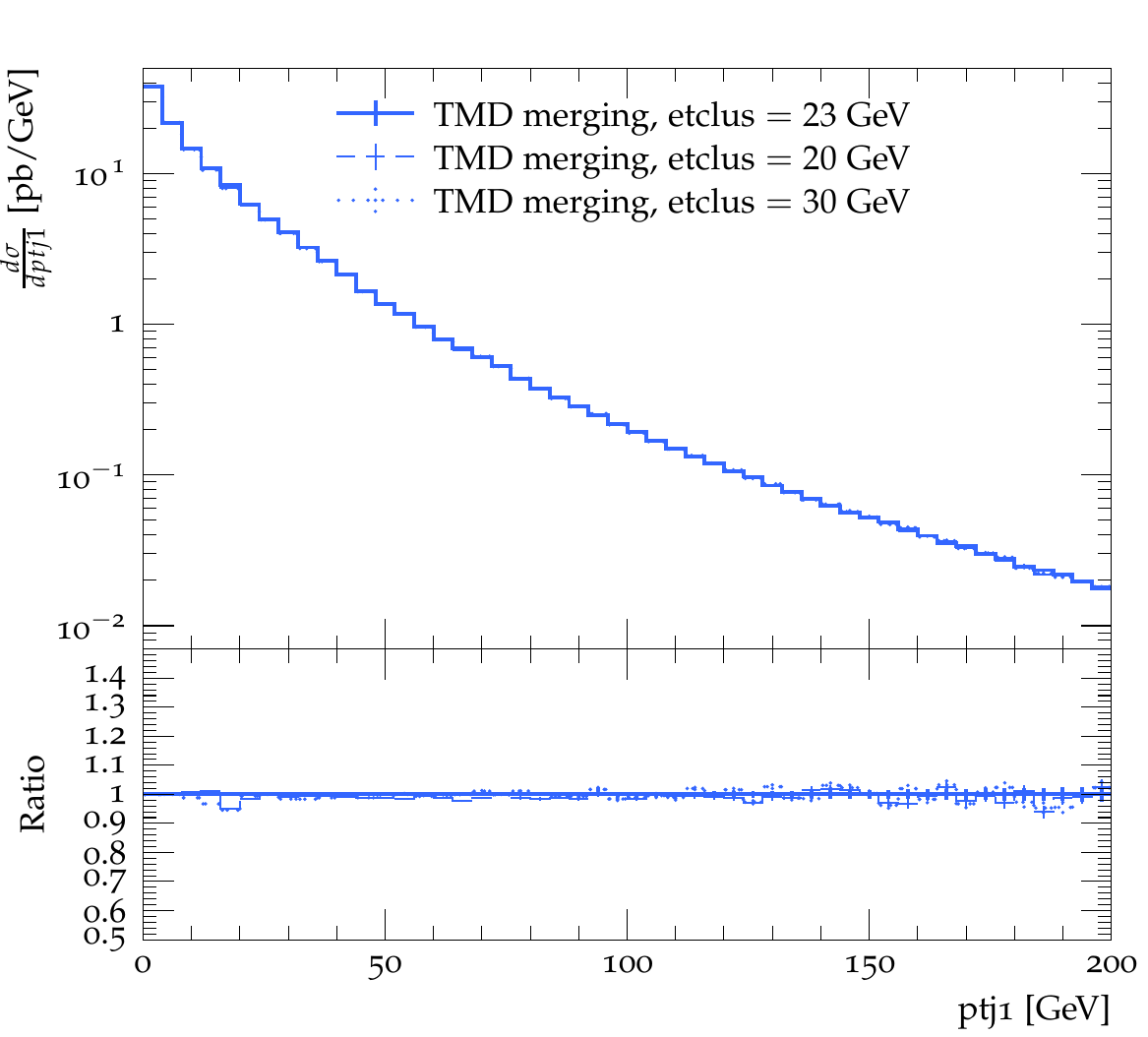} 
  \caption{Theoretical systematics studies with TMD merging in DY lepton-pair 
  production with associated jets at the LHC. (top left) 
  $\phi^\ast$ distribution of DY lepton pairs; (top right)   differential jet rate $d_{01}$; 
  (bottom) leading jet $p_T$.  In each panel, results are shown for 
  three different values of the merging scale, with the solid line giving the default setting at merging scale of 23 GeV.}
  \label{fig4-systematics}
  \end{center} 
\end{figure}

The very good agreement of the prediction with the experimental measurements in 
Fig.~\ref{fig2-multiplicities} 
 illustrates that  the number of jets which result into the lepton pair $p_T$ imbalance is well described by the TMD merging calculation. 
We observe that the agreement holds up to multiplicities much larger than the maximum number of jets (three) for 
 which the exact LO matrix-element calculation is performed. This underscores the potential benefit of the TMD evolution in better 
 describing hard and non-collinear emissions, compared to the standard collinear evolution.

In Ref.~\cite{Martinez:2021chk} we have further examined 
  the transverse momentum spectra of the associated jets.  
The comparison of the TMD merging results~\cite{Martinez:2021chk}  
with the experimental measurements~\cite{Aaboud:2017hbk} 
indicates that TMD merging  describes well not only  the number of jets 
 (as seen in Fig.~\ref{fig2-multiplicities}) but also the $p_T$ of 
the leading jet.  
Furthermore, 
one 
can compare the results of the TMD merging calculation with the results from the collinear merging calculation 
which is obtained by 
 replacing the initial-state TMD shower evolution with collinear shower evolution implemented in {\sc Pythia}6,   
while keeping the same  matrix-element and final-state shower in the two calculations.  
It is found~\cite{Martinez:2021chk} that 
clear differences emerge in 
the spectra that are most sensitive to higher-order shower emissions, such as the leading jet $p_T$ distribution in final states 
with at least 4 jets. The description of the jet $p_T$ improves thanks to TMD with respect to collinear merging 
at high multiplicities. 

\section{Theoretical systematics} 

We finally turn to  the theoretical systematics  
associated with the multi-jet merging algorithm,  and in particular the dependence of theoretical predictions 
on the merging scale. It is shown in Ref.~\cite{Martinez:2021chk} that the multi-jet rates in $Z$-boson + jets production, computed 
with TMD merging 
for different multiplicities with the phase space selection and cuts of~\cite{Aaboud:2017hbk},   have 
 variations of less than 2\% for a 10 GeV variation of the merging scale. This systematic uncertainty is significantly 
 smaller than that of standard algorithms of collinear merging, as reported in Ref.~\cite{Alwall:2007fs}, where the variation 
 of the jet multiplicity rates is found to be about 10\% when a 10 GeV change in the merging scale is considered.

Besides the effects on the rates,  in Fig.~\ref{fig4-systematics} we present results for 
differential distributions in $Z$-boson + jets events, obtained by using  
 the  TMD merging algorithm and   varying the merging scale around the default value. As in the previous calculations, the 
 default value of the merging scale is taken to be 23 GeV,   and we consider variations to 20 GeV and 30 GeV. We show 
 results for the $\phi^\ast$ distribution of lepton pairs~\cite{Aad:2015auj}, the  DJR $d_{01}$, the leading jet $p_T$. We 
 observe that the variations  in the distributions  are localized around the merging scale, and the size of the variations is  
 within 10\%.  
 
 Our  findings indicate  that the systematic uncertainties  are reduced owing to TMD merging with respect to collinear merging. 
 In general, the merging systematics reflects the mismatch between the matrix-element and parton-shower weights assigned 
 to a given final state. The larger the mismatch, the larger the uncertainty. The phase space regions that are most affected are 
 those describing final states for which the jet multiplicity can vary under small changes of the merging parameters. For instance, 
 this happens if a jet is soft or close to another hard jet.  Modeling better the emission probability for such jets by shower evolution, by treating the 
 transverse momentum recoils   through  TMD 
  distributions,  reduces the difference with the weight assigned to these events by the matrix element description, thereby 
  reducing the mismatch and the relative systematics. 

\section{Conclusion}

We have discussed implications of TMD parton evolution on multi-jet production in high-energy hadronic collisions. 
We have presented a new multi-jet merging method, TMD merging, that complements current methods with the use of TMD parton branching 
for the initial-state evolution.   

Compared to standard approaches such as MLM, we find that 
TMD merging (i)~has reduced systematic uncertainties, and (ii)~improves the 
description of higher-order emissions beyond the maximum parton multiplicity considered in the matrix element calculations.

As the TMD broadening grows with 
increasing evolution scale $\mu$ and decreasing longitudinal momentum fraction $x$, we expect the 
effects pointed out in this article  to become even more relevant  in the case of the higher  scales probed with jets at 
higher luminosity~\cite{Azzi:2019yne} and 
higher energy~\cite{Mangano:2016jyj} colliders.

\acknowledgments
We thank the organizers of the Rencontres for the invitation, and 
for the excellent online edition of the conference.

\end{document}